\documentclass{article}
\usepackage{spconf,amsmath,graphicx,psfrag}
\usepackage{bm}
\usepackage{amsfonts}
\usepackage{amsmath}
\usepackage{auto-pst-pdf}

\title{FADI-AEC: Fast Score Based Diffusion Model Guided by Far-end Signal for Acoustic Echo Cancellation}
\name{Yang Liu, Li Wan, Yun Li, Yiteng Huang, Ming Sun, James Luan, Yangyang Shi, Xin Lei}
\address{Meta Platforms, USA \\
\{yangliuai, wwanli, yunli1, yah, sunming425, jamesluan, yyshi, leixin\}@meta.com}

\begin{document}
%
\maketitle
\begin{abstract}

Despite the potential of diffusion models in speech enhancement, their deployment in Acoustic Echo Cancellation (AEC) has been restricted. In this paper, we propose DI-AEC, pioneering a diffusion-based stochastic regeneration approach dedicated to AEC. Further, we propose FADI-AEC, fast score-based diffusion AEC framework to save computational demands, making it favorable for edge devices. It stands out by running the score model once per frame, achieving a significant surge in processing efficiency. Apart from that, we introduce a novel noise generation technique where far-end signals are utilized, incorporating both far-end and near-end signals to refine the score model's accuracy. We test our proposed method on the ICASSP2023 Microsoft deep echo cancellation challenge evaluation dataset, where our method outperforms some of the end-to-end methods and other diffusion based echo cancellation methods.

\end{abstract}
\begin{keywords}
echo cancellation, diffusion model, stochastic regeneration, predictive learning
\end{keywords}

\section{Introduction}

The importance of acoustic echo cancellation in achieving high-quality speech in voice communication has led to the emergence of deep neural network (DNN) based methods, such as the deep complex convolution recurrent network (DC-CRN) \cite{hu2020dccrn}. The main challenges are artifacts, target speech distortion, and echo leakage during double-talk scenarios. To address these challenges, researchers have proposed techniques such as alignment modules \cite{ liu2023sca}, novel architecture \cite{zhang2021ft, zhao2022deep} and modified loss functions \cite{xiong2023deep}. Currently, the majority of echo cancellation models adhere to predictive methodologies, which learn a deterministic mapping from corrupted speech to clean speech targets. However, generative models, such as variational auto-encoders (VAEs) \cite{kingma2013auto}, generative adversarial networks (GANs) \cite{goodfellow2020generative}, and diffusion approaches \cite{sohl2015deep}, offer a different perspective. These models learn the target distribution and have the capability to generate multiple valid estimates, providing a richer set of possibilities for AEC task.

Among these, diffusion-based generative models have shown promising results in tasks such as noise suppression. For instance, Lu et al. \cite{lu2022conditional} proposed a novel approach that incorporates the characteristics of the observed noisy speech signal into the diffusion and reverse processes, providing a more tailored enhancement. Similarly, Joan Serrà et al. \cite{serra2022universal} introduced a multi-resolution conditioning network that employs score-based diffusion. Their model generates clean speech from noisy inputs by progressively reducing the noise in a series of steps, effectively diffusing the noise out of the signal. Lemercier et al. \cite{lemercier2022storm} applied a stochastic regeneration approach, where an estimate provided by a predictive model is used as a guide for further diffusion, refining the enhancement process.

While there have been significant advancements in the field, the potential of using diffusion models for echo cancellation has largely remained untapped. One of the primary challenges is the computational intensity of existing diffusion models, which makes their deployment in real-world production settings a daunting task. In this paper, we introduce two novel models: DI-AEC, which applies a diffusion-based stochastic regeneration method to echo cancellation, and FADI-AEC, an efficient version of DI-AEC that utilizes a fast score model. To the best of our understanding, these represent the first echo cancellation models that harness the power of diffusion. Notably, FADI-AEC addresses the computational hurdle by executing the score model only once per frame, leveraging the prior state to minimize processing time. Both models employ far-end signals to produce noise, thereby enhancing performance. By considering both far-end and near-end signals, these models enhance accuracy, generating high-quality samples.


\begin{figure}[t]
    \centering
    { 
      \small
      \psfrag{x}[l]{$x(n)$}
      \psfrag{D}[c][b]{$D_{\theta}$}
      \psfrag{G}[c][b]{$G_{\phi}$}
      \psfrag{h}[c]{$h(n)$}
      \psfrag{s}[c]{$\hat{s}(n)$}
      \psfrag{z}[c]{$\hat{s}(n)-\sigma(n)z(x(n))$}
      \psfrag{t}[c]{$\tilde{s}(n)$}
      \includegraphics[width=\columnwidth]{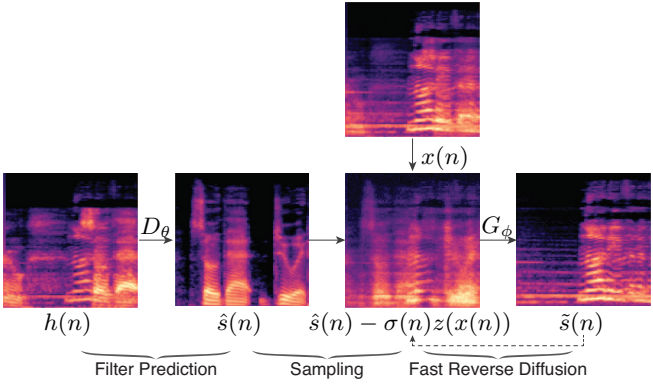}
      \vspace{-4ex}
    }
    \caption{FADI-AEC pipeline. The predictive filter is first used to generate an predicted estimate $\mathbf{\hat{\mathbf{s}}}(n)$ from mic signal $\mathbf{h}(n)$. Diffusion-based generation $G_{\phi}$ is then performed by adding Gaussian noise guided by far-end sign $\mathbf{x}(n)$ and solving the reverse diffusion SDE. The estimated near-end speech is $\mathbf{\tilde{s}}(n)$, which would be used in the score function in the next frame.}
    \label{fig:my_label}
\end{figure}

\section{Proposed method}

\subsection{Problem formulation}
In a typical AEC system, the microphone signal is denoted as $\mathbf{h}(n)$. This signal comprises two components: the near-end speech $\mathbf{s}(n)$ and the acoustic echo $\mathbf{z}(n)$. Mathematically, this relationship is expressed as:

\begin{equation}
\mathbf{h}(n) = \mathbf{s}(n) + \mathbf{z}(n),
\end{equation}
where $n$ is the time sample index. The acoustic echo $\mathbf{z}(n)$ can be understood as a time-delayed version of the far-end reference signal $\mathbf{x}(n)$. This signal has traversed the echo path and might have undergone nonlinear distortions due to the loudspeakers. The primary objective of the AEC system is to separate the near-end speech $\mathbf{s}(n)$ from the microphone signal $\mathbf{h}(n)$.

\subsection{Forward and inference through reverse sampling}
The stochastic forward process utilized in score-based diffusion models is defined by the Itô Stochastic Differential Equation (SDE) \cite{song2020score}:
\begin{equation}
\label{forward}
\mathrm{d} \mathbf{s}(n)_t = \mathrm{f}(\mathbf{s}(n)_t, t) \mathrm{d}t + \mathrm{g}(t) \mathrm{d} \mathbf{w}
\end{equation}
where $\mathbf{w}$ denotes a standard-dimensional Brownian motion, making $\mathrm{d} \mathbf{w}$ a zero-mean Gaussian random variable with variance proportional to $\mathrm{d} t$, pertinent for each Time-Frequency (T-F) bin. The functions $\mathrm{f}$ and $\mathrm{g}$ represent the drift and diffusion coefficients, respectively. The state of the process at discrete index $n$ and continuous time $t$, where $t \in [0, T]$ is given by $\mathbf{s}(n)_t$, and for clean speech, the initial condition is $\mathbf{s}(n)_0 = \mathbf{s}(n)$.


In the reverse process of score based diffusion model, the  score model is substituted into the reverse SDE as plug-in reverse SDE \cite{huang2021variational}:
\begin{equation}
\label{reverse}
    \begin{aligned}
& \mathrm{d} \mathbf{s}(n)_t= \\
& \quad\left[-\mathrm{f}\left(\mathbf{s}(n)_t, t\right)+\mathrm{g}(t)^2 \nabla_{\mathbf{s}(n)_t} \log p_t\left(\mathbf{s}(n)_t\right)\right] \mathrm{d} t+\mathrm{g}(t) \mathrm{d} \overline{\mathbf{w}}
\end{aligned}
\end{equation}
where $\mathrm{d} \overline{\mathbf{w}}$ is a d-dimensional Brownian motion for the time flowing in reverse and $\nabla_{\mathbf{s}(n)_t} \log p_t\left(\mathbf{s}(n)_t\right)$ is the score function. This equation is classified under the Ornstein-Uhlenbeck SDEs \cite{oksendal2013stochastic}.

During inference, Eq. \eqref{reverse} is evaluated using the predictor-corrector approach as informed by the score-matching network described later in \cite{song2020score}. The initial state of the process is drawn from the distribution:
\begin{equation}
\mathbf{s}(n)\tau \sim \mathcal{N}_{\mathbb{C}}\left(\mathbf{s}(n)_\tau ; \mathbf{h}(n), \mathbf{x}(n) , \sigma^2(\tau) \mathbf{I}\right)
\end{equation}
This distribution essentially represents a near-end signal $\mathbf{h}(n)$ and far-end signal $\mathbf{x}(n)$, to which Gaussian noise with a variance of $\sigma^2(\tau)$ is added.

\subsection{Score model with far-end guided noise}
Since the speech enhancement including AEC task could be considered as condition generation task,  the conditioning is integrated into the diffusion process by denoting the forward process where Eq. \eqref{forward} yields the following complex Gaussian distribution for the process state $\mathbf{s}(n)_t$, known as the perturbation kernel \cite{sarkka2019applied}, $\mathcal{N}_{\mathbb{C}}\left(\mathbf{s}(n)_t ; \boldsymbol{\mu}\left(\mathbf{s}(n)_0, \mathbf{h}(n), t\right), \sigma(t)^2 \mathbf{I}\right)$
where the mean is $\boldsymbol{\mu}$ and variance is  $\sigma(t)^2$. When performing inference, one tries to solve the reverse SDE in Eq. \eqref{reverse}. For the simple Gaussian form of the perturbation kernel $p_{0, t}\left(\mathbf{s}(n)_t | \mathbf{s}(n)_0, \mathbf{h}(n), \mathbf{x}(n) \right)$ for AEC tasl and the regularity conditions exhibited by the mean and variance, a score matching objective can be used to train the score model $ \mathbf{s}_\phi $. The score function of the perturbation kernel is $-\frac{\mathbf{s}(n)_t-\boldsymbol{\mu}\left(\mathbf{s}(n), \mathbf{h}(n)\right)}{\sigma(n)^2}$. We can reparameterize the score matching objective as follows
\begin{equation}
\label{score}
    \mathcal{J}^{(\mathrm{DSM})}(\phi)=\mathbb{E}\left[\left\|\mathbf{s}_\phi\left(\mathbf{s}(n)_t, \mathbf{h}(n), t \right)+\frac{\mathbf{z}}{\sigma(n)}\right\|_2^2\right]
\end{equation}
The clean $\mathbf{s}(n)$ and noisy $\mathbf{h}(n)$ utterances are picked in the training set and the current process state is obtained as $\mathbf{s}(n)_t =\boldsymbol{\mu} (\mathbf{s}(n), \mathbf{h}(n)) + \sigma(n) \mathbf{z}(\mathbf{x}(n))$. Note that the noise here is generated based on the aligned far-end signal calculated by \cite{liu2023sca}, not the traditional uniform random noise. This distinction arises because echo cancellation is different from noise suppression. The echo has a strong correlation with the far-end signal $\mathbf{x}(n)$. Utilizing $\mathbf{x}(n)$ can help reverse diffusion achieve stable performance with fewer frames. Therefore, we defined the far-end signal guided noise as $\mathbf{z}  \sim \mathcal{N}_{\mathbb{C}}(\mathbf{z} ; \mathbf{x}(n), \mathbf{I})$.

\subsection{Fast Score model}

Assuming that $\mathbf{s}(n)$ represents a stable recording, both  echo recording $\mathbf{z}(n)$ and noisy recording $\mathbf{h}(n)$  would also be stable. As more input data is taken into account, the filter in AEC would estimate the echo path, resulting in a reduced value of $\sigma$. We can think of the filter's convergence process as an inverse operation. The perturbation kernel could be simplified as:
\begin{equation}
\label{kernal}
p\left(\hat{\mathbf{s}}(n) | \mathbf{s}(n), \mathbf{h}(n)\right) = \mathcal{N}_{\mathbb{C}}\left(\hat{\mathbf{s}}(n) ; \boldsymbol{\mu}\left(\hat{\mathbf{s}}(n), \mathbf{h}(n)\right), \sigma(n)^2 \mathbf{I}\right)
\end{equation}
where $\hat{\mathbf{s}}(n)$ the current process state after AEC filter $D_{\theta}$. Further, The score function of the perturbation kernel is simplified as $-\frac{\hat{\mathbf{s}}(n)_t-\boldsymbol{\mu}\left(\mathbf{s}(n), \mathbf{h}(n)\right)}{\sigma(n)^2}$ and the score mathcing objective Eq. \eqref{score} is simplified as
\begin{equation}
\label{score_fast}
    \mathcal{J}^{(\mathrm{DSM})}(\phi)=\mathbb{E}\left[\left\|\hat{\mathbf{s}}_\phi\left(\mathbf{s}(n), \mathbf{h}(n), s(n-1)\right)+\frac{\mathbf{z}}{\sigma(n)}\right\|_2^2\right]
\end{equation}
where $\mathcal{J}^{(\mathrm{DSM})}$ is not related to $t$ , but rather to $s(n-1)$. This implies that the score model takes into account the enhanced signal from the previous frame. Optimization is achieved through iterative processing over time, rather than multiple score calculations within a single frame. While this approach might require more iterations to stabilize the system, it reduces the computational load for each frame. Subsequently, the corrector employs the score network's output to synchronize the generated sample with the anticipated marginal distribution, taking cues from the score network's estimate. For simplicity, we term $ G_\phi $ as the generative model. This corresponds to the reverse diffusion process solver, steered by both the plug-in SDE and the score network $ \mathbf{s}_\phi $. Thus, our concluding estimate is expressed as $ \hat{x}(n) = G_\phi(\mathbf{h}(n)) $.

\subsection{DI-AEC and FADI-AEC}
To address the computational burden of full diffusion-based models, we consider to use the stochastic regeneration model \cite{lemercier2022storm} as our baseline. By utilizing the predictive model's initial approximation, it significantly reduces the number of diffusion steps required, thereby curbing the computational demands typically seen with full diffusion models. DI-AEC and FADI-AEC have similar pipelines. The main different is DI-AEC uses the score \eqref{score} while FADI-AEC uses the fast score function \eqref{score_fast} and only calculate one time for each frame.  The whole pipeline of FADI-AEC is shown as Fig. 1. For near-end signal $\mathbf{h}(n)$, a predictive model $D_\theta$ serves as an initial predictor producing an estimate $\hat{\mathbf{s}}(n)$ defined as:
\begin{equation}
    \hat{\mathbf{s}}(n) = \mathbf{s}(n) - {\mathbf{s}(n)}^{dis} + {z}'(n)
\end{equation}
where ${\mathbf{s}(n)}^{dis}$ is the target distortion introduced by the predictive model. The residual corruption ${z}'(n)$ is what remains of the noise after being processed by the model. A diffusion-based generative model $G_\phi$ is then used to learn the distribution of the ideal residue $r_\mathbf{s}(n) = {\mathbf{s}(n)}^{dis} - {z}'(n)$, starting from the echo residue $r_\mathbf{h}(n) = \mathbf{h}(n) - D_\theta (\mathbf{h}(n)) = {\mathbf{s}(n)}^{dis} + \mathbf{z}(n) - {z}'(n)$. The resulting a priori Signal-echo rate in the starting point is very high, as for a reasonable predictor $D_\theta$, for $||\tilde{z}'(n)|| \ll ||\mathbf{z}(n)||$. The estimate $\tilde{s}(n)$ is then obtained as $G_\phi(D_\theta(\mathbf{h}(n)))$.

\subsection{Loss}
For training, we define the loss as $\mathcal{L}$ combining score matching and and a supervised regularization term —e.g. mean square error—matching the output of the initial predictor to the target speech:
\begin{equation}
\begin{aligned}
& \mathcal{L}^{(\text {StoRM })}(\theta, \phi) = \mathcal{L}^{(\mathrm{DSM})}(\theta)+\alpha \mathcal{L}^{(\text {Sup })}(\phi) \\
= & \mathbb{E}\left[\left\|\mathbf{s}_\phi\left(\hat{\mathbf{s}}(n), \mathbf{h}(n), s(n-1)\right)+\frac{\mathbf{z}}{\sigma(n)}\right\|_2^2\right]  \\ &+\alpha \mathbb{E}\left[\left\|\mathbf{s}(n)-D_\theta(\mathbf{h}(n))\right\|_2^2\right],
\end{aligned}
\end{equation}
where $\alpha$ is a balance term that we empirically set to 1.

\section{Experimentation Results}

\subsection{Data Selection and Enhancement}

\begin{table*}[]\small
\centering
\begin{tabular}{ccccccccc}
\hline
\textbf{Index} & \textbf{Model} & \textbf{Sampling} & \textbf{Reversion} & \textbf{\# Parameters} & \textbf{Latency} & \textbf{ERLE of} & \textbf{PESQ of} & \textbf{PESQ of} \\ 
& & & \textbf{Diffusion} & (M) & (ms) & \textbf{FEST} (dB) & \textbf{NEST} & \textbf{DT} \\ \hline
1  & CRN  & No                  & No                  & 3.6           & 4.04         & 67.67        & 4.41         & 2.34       \\\hline
2  & CRN  & No                  & No                  & 7.8           & 8.93         & 82.45        & 4.50         & 2.60       \\ \hline
3  & DI-AEC   & Random      & Yes                 & 6.9           & 325.00       & 92.51        & 4.87         & 3.30       \\\hline
4  & DI-AEC   & Far Guided  & Yes                 & 6.9           & 325.00       & 92.83        & 4.91         & 3.32       \\ \hline
5  & FADI-AEC & Random      & Fast Score      & 6.9           & 9.14         & 86.24        & 4.81         & 3.08       \\\hline
6  & FADI-AEC & Far Guided  & Fast Score      & 6.9           & 9.14         & 89.41        & 4.83         & 3.21      \\ \hline
\end{tabular}
\caption{Performance comparison over candidate models. We measure WB-PESQ both DT and NEST scenarios and ERLE for FEST scenario in the augmented evaluation dataset.}
\end{table*}

\begin{table}[]\small
\centering
\begin{tabular}{cccc}
\hline
\textbf{Model} & \textbf{FEST} & \textbf{DT} & \textbf{DT other} \\ \hline
ByteAudio & 4.709 & 4.770 & 4.312    \\ \hline
DI-AEC & 4.732 & 4,783 & 4.328    \\ \hline
FADI-AEC  & 4.719 & 4.781 & 4.321   \\ \hline
\end{tabular}
\caption{AECMOS comparison on ICASSP 2023 AEC Challenge blind test. }
\end{table}

For model training, we utilize a combination of synthetic data from the AEC-challenge \cite{cutler2022AEC} and our privately enhanced dataset. We ensure gender balance among speakers on both the far-end and near-end sides, resulting in 720 original conversations, each lasting 10 seconds. 

Each conversation is then enhanced considering several typical use cases. \textbf{Reverberation Time (RT60)}: We employ the image method \cite{allen1979image} to generate both steady and time-varying room impulse responses (RIR), which are used to simulate echo paths in common laptop environments. The RT60 values are selected based on probabilities of 0.6, 0.3, 0.08, and 0.02 for ranges of 50-300 ms, 300-600 ms, 600-1 s, and 1-1.5 s, respectively. \textbf{Delay}: We introduce a delay between the playback and its received echo, with probabilities of 0.05, 0.6, 0.4, and 0.05 for delay ranges of -20-0 ms, 0-200 ms, 200-400 ms, and 400-600 ms, respectively. \textbf{Signal-to-Noise Ratio (SNR)}: We simulate SNR using typical noises from the DNS-challenge \cite{dubey2022icassp}, with probabilities of 0.1, 0.1, 0.3, and 0.5 for SNR ranges of 0-10 dB, 10-20 dB, 20-30 dB, and 30-40 dB, respectively. \textbf{Signal-to-Echo Ratio (SER)}: We simulate SER with probabilities of 0.1, 0.5, 0.3, and 0.1 for SER ranges of -10-0 dB, 0-10 dB, 10-30 dB, and 30-40 dB, respectively. \textbf{Non-linearity}: We model non-linearity using either an arc-tangent function to mimic gain saturation or a polynomial function as demonstrated by \cite{zhang2022lcsm}. \textbf{Time-Variant Delay/RIR Changes}: We introduce random cuts or additions of speech and silence segments, ranging from 10 to 200 ms, to either near-end or far-end signals with a probability of 0-10\%. Time-variant RIR changes are also incorporated when generating the echo component. Each enhanced conversation is then converted into far-end single talk (FEST), near-end single talk (NEST), and double talk (DT) scenarios. This enhancement process results in a total of 720K enhanced conversations, approximately equivalent to 2,000 hours of conversation.

\subsection{Ablation study}
Table 1 presents the performance of various acoustic echo cancellation (AEC) models based on different sampling methods and the presence of reversion diffusion. The models are evaluated on parameters such as ERLE of FEST, PESQ of NEST, and PESQ of DT.

Initially, we can see that the CRN model \cite{hu2020dccrn} without any sampling or reversion diffusion presents two different parameter sizes. Specifically, the larger model with large latency shows superior performance in both ERLE of FEST and PESQ of NEST. This suggests that, given the same architectural characteristics, increasing the model size can lead to better AEC performance. The DI-AEC models (id 3 and 4) introduce reversion diffusion and use distinct sampling methods - Random Sampling and Far Guided Sampling, respectively. These models exhibit a notably high latency of 325.00ms. Despite the higher latency, the DI-AEC models surpass the CRN models in all the performance metrics. Shifting to FADI-AEC models (id 5 and 6), we observe that they combine a modified reversion approach, termed Fast Reversion, with sampling techniques. These models manage to significantly reduce the latency to 9.14ms. Performance-wise, the FADI-AEC with Far Guided Sampling marginally outperforms its Random Sampling counterpart, especially in terms of ERLE of FEST and PESQ of DT. To conclude, Upon applying the DI-AEC model, the DT PESQ improved by 27.7\%, rising from 2.6 to 3.32. When implementing FADI-AEC, the DT PESQ decreased to 3.08, representing a decrease of 7.2\%. However, a notable advantage of FADI-AEC is its reduced latency, which is only 2.8\% of DI-AEC's latency, comparing 9.14 to 325.

\subsection{Comparison with the state-of-the-art methods}

We use AECMOS, a non-intrusive model-based metric from the AEC challenge, to compare our proposed method against the established baselines ByteAudio \cite{zhang2023two} and DI-AEC. ByteAudio employs a Two-step Band-split Neural Network (TBNN) methodology for full-band acoustic echo cancellation, achieving the highest performance in the AEC 2023 challenges, second only to the host. ByteAudio represents an advanced end-to-end AEC network solution. Meanwhile, DI-AEC stands as another leading AEC technique optimized for various real-world acoustic echo cancellation scenarios. As shown in Table 2, FADI-AEC slightly outperforms both ByteAudio and DI-AEC across all three scenarios. Specifically, FADI-AEC attains performance scores of 4.719, 4.781, and 4.321 for the FEST, DT, and DT other scenarios respectively. All three methods demonstrate remarkable performance on the AEC task, but FADI-AEC holds a slight edge across the considered scenarios. These results underscore the importance of selecting the appropriate AEC method for specific tasks and scenarios, and also highlight the progress and potential of contemporary AEC technologies.

\section{Conclusion}

We proposed DI and FADI, two novel score-based diffusion models specifically designed for acoustic echo cancellation. The research highlights the diffusion-based stochastic regeneration model can improve AEC model performance. To address the computational cost challenges posed by DI-AEC, particularly for edge devices, FADI-AEC significantly improves processing efficiency by running the score model only once per frame.  Furthermore, FADI-AEC and DI-AEC can use a pioneering noise generation technique with far-end signals, integrating both far-end and near-end signals to enhance the precision of the score model.
This unique application of diffusion models offers a powerful and efficient approach to echo cancellation, showcasing strong performance even in challenging conditions. In future work, the fast score model proposed in FADI could be applied in the other speech enhancement task such as noise suppression and dereverberation.

\bibliographystyle{IEEEbib}
\bibliography{strings,refs}

\end{document}